\documentstyle[11pt]{article}
\input{amssym.def}
\def\pa{\partial}
\def\Im{\mathop{\rm Im}\nolimits}

\begin{document}
\begin{center}
{\Large {\bf Time Dependent Supersymmetry in Quantum Mechanics}}
\footnote{Talk given at the 7-th Lomonosov
Conference "Problems of Fundamental Physics", 24-30 August, 1995,
see proceedings book (with the minor corrections) Moscow, 1997, p. 54-61.}

\vspace{5mm}
{\bf Vladislav G. Bagrov}\footnote{E-mail: bagrov@phys.tsu.tomsk.su}
and {\bf Boris F. Samsonov}\footnote{E-mail: samsonov@phys.tsu.tomsk.su}

\vspace{2mm}
Tomsk State University, 36 Lenin Ave., Tomsk 634050, Russia\\
High Current Electronics Institute, 4 Akademichesky Ave., Tomsk 634055,
Russia
\end{center}

%\vspace{2mm}
%\maketitle
\begin{abstract}
The well-known supersymmetric constructions such as Witten's
supersymmetric quantum mechanics, Spiridonov--Rubakov parasupersymmetric
quantum mechanics, and higher-derivative SUSY of Andrianov et al.
are extended to the nonstationary Schr\"odinger equation. All these
constructions are based on the time-dependent Darboux transformation.
The superalgebra over the conventional Lie algebra is constructed.
Examples of time-dependent exactly solvable potentials are given.
\end{abstract}

\vspace{2mm}
{\bf 1.} The supersymmetric quantum mechanics is in a permanent intensive
development since the Witten papers \cite{1}. One can cite the
$N$-extended supersymmetric quantum mechanics \cite{2},
parasupersymmetric quantum mechanics \cite{3}, and higher order
derivative supersymmetric quantum mechanics \cite{4}. The field of
supersymmetric quantum mechanics is recently reviewed in \cite{5}. We
want to point out that all above mentioned constructions are valid for the
time independent Hamiltonians and if one restrict oneself by the
stationary solutions of the Schr\"odinger equation. Hence, these
constructions can be referred to the stationary supersymmetric quantum
mechanics and the nonstationary one needs to be developed. We hope that
this report gives a stimulus for the further developments in this domain.

{\bf 2.} The nonstationary supersymmetric quantum mechanics is based on the
nonstationary Darboux transformation \cite{6} just in the same way as
the stationary one \cite{1,7} is based on the conventional Darboux
transformation \cite{8}.

Let us consider two time-dependent Schr\"odinger equations
\begin{eqnarray}
&(i\pa_t-H_0)\psi(x,t)=0,\quad \pa_t=\pa/\pa t,\quad
H_0=-\pa^2_x-V_0(x,t), \quad \pa^2_x=\pa_x\pa_x,\\
&(i\pa_t-H_1)\varphi(x,t)=0,\qquad H_1=-\pa^2_x-V_1(x,t), \qquad
x\in R, \quad t\in{\Bbb R}^1.
\end{eqnarray}
Here $-V_0(x,t)$ is a potential energy and $R=[a,b]$ is the interval
for  variable $x$ which can be both finite and infinite. If the
Schr\"odinger operators for Eqs. (1) and (2) are related by the
intertwining relation
\begin{equation}
L(i\pa_t-H_0)=(i\pa_t-H_1)L,
\end{equation}
where $L$ is a linear operator, called {\it transformation operator}, the
functions $\psi$ and $\varphi$ are related as follows: $\varphi=L\psi$
if $L\psi\ne0$.

If $(i\pa_t-H_0)$ and $(i\pa_t-H_1)$ are self-adjoint (in the sense of
some scalar product) the equation (3) implies
\begin{equation}
L^+(i\pa_t-H_1)=(i\pa_t-H_0)L^+,
\end{equation}
where the superscript plus sign $(^+)$ is used to denote the operator
adjoint to $L$, and Eqs. (1) and (2) become ``peer''. It follows from
Eqs. (3) and (4) that $s_0=L^+L$ commutes with $(i\pa_t-H_0)$ and
$s_1=LL^+$ commutes with $(i\pa_t-H_1)$ and consequently $s_0$ is a
symmetry operator for the initial equation (1) and $s_1$ is a symmetry
operator for the final one (2).

The constructions such as in Eq. (3) are well-known in mathematics and
are intensively investigated since the Delsart's paper \cite{9}. The
most significant results obtained with the help of the transformation
operators concern the backscattering problem in quantum mechanics
\cite{10} and its application for the solving of the nonlinear
equations \cite{11}.

{\bf 3.} We now assume operator $L$ to be  a differential of the first degree in
$\pa_x$ with smooth coefficients depending on both variables
$x$ and $t$. We should not include in $L$ the derivative $\pa_t$ since
it, being found from equation (1), transforms $L$ into the second-order
differential operator. In this case the operator $L$ and the real
potential difference $A(x,t)=V_1(x,t)-V_0(x,t)$ are completely defined
by a function $u(x,t)$ called {\it transformation function} \cite{6}:
\begin{eqnarray}
& L=L_1(-u_x/u+\pa_x),\\
& L_1=L_1(t)=\exp\big(2\int dt\,\Im(\log u)_{xx}\big),\\
& A=(\log|u|^2)_{xx}.
\end{eqnarray}
To obtain a real potential difference we should impose the reality condition \cite{6} on the function $u$
\begin{equation}
(\log u/u^*)_{xxx}=0,
\end{equation}
where the asterisk implies the complex conjugation.

In the majority of cases of physical interest we can introduce the
Hilbert space structure $L^2_0(R)$ in the space of solutions of the
equation (1) with the scalar product appropriately defined. Symmetry
operator $s_0=L^+L$, being symmetric, may be extended up to self-adjoint in the appropriate Hilbert space and can have either discrete
spectrum or continuous one. Since $Lu=0$ (see Eq.
(5)), the function $u$ is an eigen function of this operator corresponding to zero
eigen value. Hence, $u$ is one of the eigen functions of
operator $L^+L=h-\alpha$. In general, $h$ is a self-adjoint integral
of motion for the initial Schr\"odinger equation (1) which in
particular case (if $V_0$ does not depend on $t$) may be equal to the
Hamiltonian $H_0$.

With the help of the transformation operator $L$, just in the same way
that in the conventional supersymmetric quantum mechanics \cite{1,7}, we
can construct the supercharge operators
\begin{equation}
Q=\left(\begin{array}{cc} 0 & 0\\ L & 0\end{array}\right)=(Q^+)^{\dag}
\end{equation}
acting on two-component wavefunctions $\Psi(x,t)=\left(
\begin{array}{c} \psi(x,t)\\ L\psi(x,t)\end{array}\right)$.
Two Schr\"o\-din\-ger equations (1) and (2) may now be rewritten in supersymmetric form
\begin{equation}
(iI\pa_t-{\cal H})\Psi(x,t)=0,
\end{equation}
where $I$ is $2\times2$ identity matrix and ${\cal H} =\left(
\begin{array}{cc} H_0 & 0\\ 0 & H_1\end{array}\right)$ is a
superhamiltonian.
Since $s_0=L^+L$ and $s_1=LL^+$ are symmetry
operators for equations (1) and (2) respectively, the superoperator
$S=\left(\begin{array}{cc} L^+L & 0\\ 0 & LL^+\end{array}\right)$ is
the symmetry operator for the equation (10). The operators $Q$, $Q^+$,
and $S$ form a well-known superalgebra \cite{1,7}. There is a single
difference, namely, instead of the Hamiltonians we see other
integrals of motion of the equation (1) in its construction. When $h$ is the
initial Hamiltonian
above constructions coincide with known ones. This is the
reason to call the transformation  (5), (6), {\it time-dependent
Darboux transformation} \cite{6}.

4. With the help of the other eigen functions of operator $h$ we can
perform the chain of Darboux transformations and construct the
parasuperalgebra in full analogy with papers \cite{3}. If in this chain
we eliminate all intermediate operators $h$ and express the final
operator $L$ in terms of particular solutions
$u_i$ of the initial equation (1), we obtain higher-derivative nonstationary
quantum mechanics similar to the stationary one \cite{4}. In this
case
\begin{equation}
L\equiv L^{(N)}=L_N(t)W^{-1}(u_1,\dots,u_N)\left|\begin{array}{cccc}
u_1 & u_2 & \dots & 1\\ u_{1x} & u_{2x} & \dots & \pa_x\\
\vdots \\ u_{1x}^{(N)} & u_{2x}^{(N)} & \dots & \pa_x^N\end{array}
\right|
\end{equation}
where $W$ stands for the conventional symbol of the Wronskian of the
functions \\ $u_1,\dots,u_N$. For the real function $L_N(t)$ we have
\begin{equation}
L_N(t)=\exp \Big\{2\int dt\,\Im[\log W(u_1,\dots,u_N)]_{xx}\Big\}.
\end{equation}
The reality condition (8) takes now the form
\begin{equation}
\Big[\log\frac{W(u_1,\dots,u_N)}{W^*(u_1,\dots,u_N)}\Big]_{xxx}=0.
\end{equation}
For the potential difference we obtain
\begin{equation}
A_N=\big(\log|W(u_1,\dots,u_N)|^2\big)_{xx}.
\end{equation}
We can recognise in formula (11) the generalisation of the known
Crum-Krein formula \cite{12,13} to the nonstationary case. Note
that the condition (13) is more feeble than the reality condition (8)
imposed on every function $u_i$. Thus, we can construct the
higher-derivative supersymmetry with the self-adjoint final Hamiltonian
even if the intermediate Hamiltonians are not self-adjoint (so-called
irreducible case described for stationary equation in Ref. 4). The basic
relation for the time-dependent polynomial supersymmetric quantum
mechanics is the following factorisation properties
\begin{equation}
L^+L=\prod_{i=1}^N(h_0-C_i), \qquad LL^+=\prod_{i=1}^N(h_1-C_i)
\end{equation}
first obtained for the stationary case in Ref. 4. The $C_i$
in Eqs. (15) are eigen values corresponding to the eigen functions $u_i$
of the operator $h_0$.

{\bf 5.} To obtain a regular potential difference by the Darboux
transformation (5)--(7) the transformation function $u$ should be
nodeless. In the space $L_0^2(R)$ a single nodeless eigen function of
the operator $h$ exists (if it has a discrete spectrum).
This function is the ground state function. Beyond the
space $L^2_0(R)$ there are many nodeless eigen functions of the
operator $h$ suitable for use as the transformation
functions. They should have eigen values
$\alpha<\varepsilon_0$ ($\varepsilon_0$ being the lowest eigenvalue
of $h$). In this case the discrete
spectra of the symmetry operators $h=L^+L+\alpha$ and $\bar h=LL^++
\alpha$ differ by a single level and we have unbroken supersymmetry
\cite{1}. Every bounded state of the superoperator $S$, except for its
ground state, is double degenerate.

We will now describe unexpected
peculiarities in the breakdown of the supersymmetry in the
higher-derivative supersymmetric quantum mechanics. These peculiarities
(as far as we know) are not discussed in the available literature.

The Darboux transformation (5,6) being performed with the discrete
spectrum function $u_n(x,t)$ of the operator $h$ [it has $n$
zeros in the interval $(a,b)$] gives the potential difference with $n$ poles.
Solutions obtained with the help of the transformation operator (5)
are not square integrable functions
in $[a,b]$. Nevertheless, the second transformation with the transformation
function $u_{n+1}(x,t)$, having $n+1$ zeros in $(a,b)$ removes all
singularities and the transformation operator of the second degree
$L^{(2)}=L_2L_1$, where $L_{1,2}$ are the first order Darboux transformation
operators, is well defined. This fact reflects the known property of the
Wronskian constructed from the functions $u_{k_i}$ belonging to the space $L^2(R)$
 \cite{13}: the Wronskian $W(u_{k_1},\dots,u_{k_N})$
conserves its sign if for all $k=0,1,2,\dots$ the inequality $(k-k_1)
(k-k_2)\cdots (k-k_N)\ge 0$ holds. In particular, the functions
$u_{k_i}$ may be two by two juxtaposed functions. The discrete eigenvalues
$\alpha_{k_i}$ of the operator $h$ are absent in the spectrum of transformed operator
$\bar H$. This signifies that the ground state level of the
superoperator $S$ is two-fold degenerate and the excited states
constructed with the help of the functions $u_{k_i}$ are nondegenerate.
Furthermore, these states are annihilated by the operators $Q$ and
$Q^+$ whereas the ground state is annihilated only by the one of
these operators. It should be noted that this property remains valid
for the stationary case, i.e., in the conventional supersymmetric quantum
mechanics.

{\bf 6.} Differential symmetry operators for the stationary Schr\"odinger
equation are Hamiltonian and its polynomial functions. Symmetry algebra of
the nonstationary Schr\"odinger equation is more
rich than for the stationary case. We can use this algebra to
construct a superalgebra. For this purpose we should define
an operator inverse to $L$.

The equation (5) implies $Lu=0$. Choose the transformation
function $u$ such that the absolute value of $u^{-1}(x,t)$ is
square-integrable in the interval $R$ and the condition (8) is satisfied.
Then for every $\psi\in L^2_0(R)$ we have $\varphi=L\psi\in L^2_1(R)$,
but the set $L^2_{11}(R)=\{\varphi: \varphi=L\psi,\,\psi\in L^2_0(R)\}$
does not span the whole space $L^2_1(R)$. The function $\varphi_0(x,t)=
[L_1(t)u^*(x,t)]^{-1}\in L^2_1(R)$ \cite{6} can not be obtain by the
action of the operator $L$ on any $\psi\in L^2_0(R)$. If we designate
by $L^2_{10}(R)$ the linear hull of the function $\varphi_0$ then
$L^2_1(R)=L^2_{10}(R)\oplus L^2_{11}(R)$.

Choose as the transformation function the function $v=\varphi_0$ and
define the following integral operator acting from $L^2_{11}(R)$ to
$L^2_0(R)$
\begin{equation}
M\varphi(x,t)=[L_1(t)v^*(x,t)]^{-1}\int_a^x v^*(y,t)\varphi(y,t)dy.
\end{equation}
The straightforward calculation persuades that $LM\varphi=\varphi$ for
all $\varphi\in L^2_{11}(R)$ and the condition $v\in L^2_{10}(R)$
implies $ML\psi=\psi$ for all $\psi\in L^2_0(R)$ operator $M$, hence,
is inverse to $L: M=L^{-1}$, and we have one-to-one correspondence
between the spaces $L^2_0(R)$ and $L^2_{11}(R)$.

If in the space $L^2_0(R)$ the symmetry operators $g_i$ forming
$n$-dimensional Lie algebra $G$ with the structural constants
$f^l_{ij}: [g_i,g_j]=f^l_{ij}g_l$ are defined and this space is
invariant under the action of these operators then in the space
$L^2_{11}(R)$ we can define the operators $\bar g_i=Lg_iM$ and this
space will be invariant under the action of all $\bar g_i$. Furthermore
these operators form a basis for the $n$-dimensional Lie algebra $\bar
G$ with the same structural constants $f^l_{ij}$.
Let $T_0$ be the space of two-component wave functions
$\Psi(x,t)$ with the basis $\Psi_+(x,t)=\psi(x,t)e_+$ and
$\Psi_-(x,t)=L\psi(x,t)e_-$, $\psi\in L^2_0(R)$, and $e_+={1 \choose
0}$, $e_-={0 \choose 1}$. In the space $T_0$ we can
define operators
\begin{equation}
G_i=\left(\begin{array}{cc} g_i & 0\\ 0 & \bar g_i\end{array}\right)
\end{equation}
which form a Lie algebra isomorphic to $G$. The operators $G_i$ are
symmetry operators for the supersymmetric equation (10). Besides the
operators $G_i$ in the space $T_0$ the following operators can be
defined: $P_0=L\sigma^-$, $Q_i=g_iL^{-1}\sigma^+$ where
$\sigma^-=\left(\begin{array}{cc} 0 & 0 \\ 1 & 0\end{array}\right)$,
$\sigma^+=\left(\begin{array}{cc} 0 & 1\\ 0 & 0\end{array}\right)$.
These operators are evidently nilpotent: $P_0^2=0$, $Q_i^2=0$, and
$\{Q_i,Q_j\}\equiv Q_iQ_j+Q_jQ_i=0$. Furthermore, we can find by the
direct calculations that $\{P_0,Q_i\}=G_i$ and the generalised Jacobi
identities are fulfilled. The operators $G_i,Q_i,P_0$, $i=1,2,\dots,n$,
hence, form a basis for $2n+1$-dimensional Lie superalgebra $sG$.

Note that since $h=L^+L+\alpha\in G$ we have for the operator $S$
introduced in sec. 3, $S\in sG$ and $Q$, $Q^+\in sG$.

{\bf 7. } Examples. Consider first the simplest case of a free particle:
$v_0(x,t)=0$. Choose the following solutions of the initial
Schr\"odinger equation (1) \cite{14}:
\begin{eqnarray}
&\psi_\lambda(x,t)=(1+t^2)^{-1/4}\exp[ix^2t/(4+4t^2)+i\lambda \arctan t]
Q_\lambda(z), \\
&z=x(1+t^2)^{-1/2},\nonumber
\end{eqnarray}
where $Q_\lambda(z)$ is the parabolic cylinder functions satisfying the
equation $Q''_\lambda(z)-(z^2/4+\lambda)Q_\lambda(z)=0$ with $\lambda$
being an arbitrary parameter (a separation constant). For
$\lambda=n+1/2$, $n=0,1,2,\dots $, the functions $Q_\lambda(z)$ are expressed
via the Hermite polynomials $Q_{n+1/2}(z)=\exp(z^2/4)H_n(iz/\sqrt{2})$.
The reality condition (8) is satisfied for all real $\lambda$.
Functions (18) being nodeless for $\lambda=n+1/2$ and for even $n$ are
suitable for use as transformation functions. Formula (7) gives the new
potential
\[ v_1^{(2k)}(x,t)=-(1+t^2)^{-1}\Big(1+4k(2k-1)\frac{q_{2k-2}(z)}{q_{2k}(z)}-
8k^2\Big(\frac{q_{2k-1}(z)}{q_{2k}(z)}\Big)^2\Big),  \]
%$$z=x(1+t^2)^{-1/2}$$
where $q_k(z)=(-i)^kHe_k(iz)$, $He_k(z)=2^{-k/2}H_k(z/\sqrt{2})$.

The same functions for odd $n$ are nodeless in the semiaxis
$(0,\infty)$ and with their help we obtain the following time-dependent
exactly solvable potential
\begin{eqnarray*}
& v_1^{(2k+1)}=-(1+t^2)^{-1}\Big(1+4k(2k+1)\displaystyle\frac{q_{2k-1}(z)}
{q_{2k+1}(z)}-2(2k+1)^2\Big(\frac{q_{2k}(z)}{q_{2k+1}(z)}\Big)^2\Big),\\
& x=z\sqrt{1+t^2}\in (0,\infty).
\end{eqnarray*}

The functions (18) for $\lambda=-n-1/2$ form a discrete basis set in
$L^2_0({\Bbb R}^1)$. The double Darboux transformation with juxtaposed
functions $u_n=\psi_{-n-1/2}$ and $u_{n+1}$ gives a regular potential
of the form
\begin{eqnarray*}
& v_2^{(n,n+1)}(x,t)=-2(1+t^2)\big[J''_n(z)/J_n(z)-\big(J'_n(z)/J_n(z)\big)^2
-1\big],\\
& J_k(z)=\Gamma(k+1)\sum\limits_{s=0}^k\Gamma^{-1}(s+1)He^2_s(z)=kJ_{k-1}(z)+
He^2_k(z),\\
& J_0(z)=1, \qquad J_1(z)=z^2+1, \qquad J_2(z)=z^4+3, \dots ~.
\end{eqnarray*}
These are the potentials which correspond to the supersymmetric model with
two-fold degenerate eigenvalue of the superoperator $S$ except for the eigenvalues corresponding to the functions $u_n$ and $u_{n+1}$. When
$n>0$ the ground state of $S$ is two-fold degenerate and nondegenerate eigenvalues are situated in the middle of the spectrum of $S$

It is not difficult to establish that the Wronskian constructed from two functions (18)
with $\lambda_1=m+1/2$ and $\lambda_2=l+1/2$ for $m=0,2,4,\dots$ and
$l=m+1,m+3,\dots$ is nodeless and therefore these functions are
suitable for double Darboux transformation. This gives the following
exactly solvable potential
\begin{eqnarray*}
& v_2^{(m,l)}(x,t)=-2(1+t^2)^{-1}\big(1+d^2\log f_{ml}(z)/dz^2\big),\\
& f_{ml}(z)=q_m(z)q_{l+1}(z)-q_l(z)q_{m+1}(z).
\end{eqnarray*}

We will cite one example for harmonic oscillator potential as well:
$H_0=-\pa^2_x+\omega^2x^2$, $H_0\psi_n=(2n+1)\Psi_n$, $\psi_n=
H_n(\sqrt{\omega}x)\exp\big(-i\omega(2n+1)t-\omega x^2/2\big)$,
$n=0,1,2,\dots $. If we choose the following nonstationary solution of
the initial Schr\"odinger equation (1) as the transformation function:
\begin
%{equation}
{eqnarray*}
&u(x,t)=\sin^{-1/2}(2\omega t)\cos h(\lambda x/\sin 2\omega
t)\times  \\
&\exp[i(\omega x^2-\lambda^2/\omega)\cot(2\omega t)/2] \notin
L_0^2({\Bbb R}^1),\ \ \lambda\in{\Bbb R}^1,
\end{eqnarray*}
we obtain the nonstationary anharmonic potential of the form:
$v_1(x,t)=\omega^2x^2-2\lambda^2\sin^{-1/2}(2\omega t){\rm sech}^2
\big(\lambda x/\sin 2\omega t)\big)$.

First author was supported by the Russian Foundation for Fundamental
Research.


\begin{thebibliography}{nn}
\bibitem{1} E. Witten, Nucl. Phys., {\bf B 188} (1981) 513; ibid., {\bf
B 202} (1982) 253.
\bibitem{2} A.I. Pashnev, Sov. J. Theor. Math. Phys., {\bf 60} (1986)
311; V.P. Berezovoy and A.I. Pashnev, Sov. J. Theor. Math. Phys., {\bf
78} (1989) 289; V.P. Berezovoy and A.I. Pashnev, Z. Phys. {\bf 51}
(1991) 525.
\bibitem{3} V.A. Rubakov and V.P. Spiridonov, Mod. Phys. Lett. {\bf
A3} (1988) 1337; A.A. Andrianov and M.V. Ioffe, Phys. Lett. {\bf B 255}
(1991) 543; J. Beckers, N. Debergh, and A.G. Nikitin, Mod. Phys. Lett.
{\bf 7} (1992) 1609.
\bibitem{4} A.A. Andrianov, M.V. Ioffe, and V.P. Spiridonov, Phys.
Lett. {\bf A 174} (1993) 273; A.A. Andrianov, M.V. Ioffe, and D.N.
Nishnianidze, {\it Polynomial SUSY in Quantum Mechanics and Second
Derivative Darboux Transformation}, preprint SPbU-IP-94-05 (1994);
A.A. Andrianov, F. Canata, J.-P. Dedonder, and M.V. Ioffe, {\it Second
Order Derivative Supersymmetry and Scattering Problem}, preprint
SPbU-IP-94-03 (1994).
\bibitem{5} F. Cooper, A. Khare, and U. Sukhtame, Phys. Rep. {\bf 251}
(1995) 267.
\bibitem{6} V.G. Bagrov, B.F. Samsonov, and L.A. Shekojan, Izv. Vyssh.
Uchebn. Zaved. Fizika No 7 (1995) 59; V.G. Bagrov and B.F. Samsonov,
Phys. Lett. A (in press).
\bibitem{7} A.A. Andrianov, N.V. Borisov, M.V. Ioffe, and I.M. Eides,
Sov. J. Theor. Math. Phys. {\bf 61} (1984) 17; C.V. Sukumar, J. Phys.
{\bf A 18} (1985) 2917; ibid., 2937; V.G. Bagrov and B.F. Samsonov, Sov.
J. Theor. Math. Phys. {\bf 104} (1995) 356.
\bibitem{8} G. Darboux, Compt. Rend. Acad. Sci. {\bf 94} (1882) 1456;
G.~Darboux, {\it Le\c{c}ons sur la theorie generale des surfaces et les
application geometrique du calcul infinitesimal}. Deuxieme partie,
Paris, Gauthier-Villars et fils (1889).
\bibitem{9} J. Delsart, J. Math. Pures et Appl. {\bf 17} (1938) 213; J.
Delsart and J.L. Lions, Comment. Math. Helv. {\bf 32} (1957) 113.
\bibitem{10} Z.S. Agranovich and V.A. Marchenko, {\it Backscattering
Problem}, Kharkov (1969); B.M. Levitan, {\it Inverse Sturm-Liouville
Problems}, Nauka, Moscow (1984); L.D. Faddeev, Usp. Mat. Nauk {\bf 105}
(1959) 57.
\bibitem{11} V.B. Matveev and M.A. Salle,
Darboux transformations and solitons, Berlin, Springer (1991);
 F. Calogero and A. Degasperis, {\it Spectral Transform and
Solitons}, Amsterdam--New York--Oxford; R.K. Dodd, J.C. Eilbeck, J.D.
Gibbon, and H.C. Morris, {\it Solitons and Nonlinear Wave Equations},
Academic Press; V.E. Zakharov, {\it Backscattering Method, Solitons}.
Ed. by R. Bulla and F.M. Kodri, Moscow (1983) 270.
\bibitem{12} M. Crum, Quart. J. Math. {\bf 6} (1955) 263.
\bibitem{13} M.G. Krein, Dokl. Akad. Nauk SSSR {\bf 113} (1957) 970.
\bibitem{14} W. Miller Jr., {\it Symmetry and Separation of Variables}.
Massachusetts (1977).
\end{thebibliography}
\end{document}